\documentclass[12pt]{article}
\usepackage{a4}
\usepackage{graphicx}
\newcommand{\be}{\begin{equation}}
\newcommand{\ee}{\end{equation}}
\newcommand{\ba}{\begin{eqnarray}}
\newcommand{\ea}{\end{eqnarray}}
\newcommand{\dsp}{\displaystyle}
\newcommand{\order}{\ensuremath{{\cal O}}}
\newcommand{\lag}{\ensuremath{{\cal L}}}
\newcommand{\im}{\mbox{Im}}
\newcommand{\re}{\mbox{Re}}
\newcommand{\disc}{\mbox{disc}}

\begin{document}
\begin{titlepage}
\begin{flushright}
BUTP-2002/03\\
LU TP 02-05\\
February 2002
\end{flushright}
\vfill
\begin{center}
{\large\bf Eta decays at and beyond $\mathbf{p^4}$ in Chiral Perturbation
Theory}\footnote{Talks presented at the Workshop on Eta Physics,
\AA{}ngstr\"om laboratory, Uppsala, October 25-27, 2001; published in the
proceedings (Physica Scripta Topical Issues T99 (2002) 34).}
\vfill
{\bf Johan Bijnens$^a$ and J\"urg Gasser$^b$}\\[0.5cm]
$^a$Department of Theoretical Physics, Lund University\\
S\"olvegatan 14A, S-22362 Lund, Sweden\\[0.5cm]
$^b$Institute of Theoretical Physics, University of Bern\\
Sidlerstrasse 5, CH-3012 Bern, Switzerland
\end{center}
\vfill
                     \\14.40.Aq      (pi  , K, and eta  mesons),
                     12.15.Ff  (Quark and lepton masses and mixing)
\begin{abstract}
An overview is given of Chiral Perturbation Theory and various applications
to eta decays. In particular, the main decay $\eta\to3\pi$ is discussed
at the one-loop level, and estimates of the higher order corrections
are given.
The importance of $p^6$ and higher order effects in double Dalitz and
$\eta\to\pi^0\gamma\gamma$ decays is pointed out.
In all these cases the need for new experimental results is stressed.
\end{abstract}
\vfill
\end{titlepage}
\clearpage
\setcounter{page}{2}
\tableofcontents
\clearpage

\section{Introduction}

Eta decays at high precision, both in measuring rare decays and in precisely
determining kinematical distributions in the more common decays, have
a high potential to teach us about various aspects of the strong interaction.
In the view of the new experimental results to be expected from WASA
at CELSIUS in Uppsala and KLOE at DAPHNE in Frascati
and the ongoing work of the Crystal Ball,
an overview of
the present situation as regards some of the main eta decays is appropriate.
The importance of these decays and measurements can be judged from the fact
that they were the main focus of several theoretical talks presented at
this workshop, in addition to the ones by the present authors there
is also the contribution by Holstein\cite{Holstein} and by
Ametller\cite{Ametller}.

In this review we concentrate on chiral symmetry aspects for
the main strong decay $\eta\to3\pi$ and how it can be used to extract
information on quark mass ratios. We will show how precise measurements
of the slope parameters will lead to more refined theoretical predictions
for the decay rate as a function of the quark mass ratio
\be
\frac{m_s-\hat m}{m_d-m_u}\,.
\ee
 A discussion of the decay $\eta\to3\pi$ in  Chiral 
Perturbation Theory (CHPT) and the use of 
dispersion theory to improve on the one-loop prediction
 is the main focus of this short review.  In addition,
a short overview is given of Dalitz ($\eta\to\gamma l^+l^-$)
and double Dalitz ($\eta\to l^+l^- l'^+l'^-$) decays, as well as of
 the process  $\eta\to \pi^0\gamma\gamma$.
 We indicate  why these are important 
quantities to be measured.

We first give a very short overview of chiral symmetry and its incorporation
using effective field theory methods - known as Chiral Perturbation
Theory - in Sect. \ref{chptintro}. The decay $\eta\to3\pi$ is then discussed
at the purely perturbative level in CHPT at order $p^4$ in Sect. \ref{p4eta}
and the estimates of higher orders using dispersion relations in Sect.
\ref{dispeta}. Section \ref{doubledalitz} is devoted to the Dalitz and double
Dalitz decays,
 and the
decay $\eta\to\pi^0\gamma\gamma$ is the subject of
section \ref{etapigg}. Our conclusions are presented in section 
\ref{conclusions}.

\section{Basics of CHPT}
\label{chptintro}

We cannot calculate all things we are interested in in the strong interaction
directly from QCD. E.g. scattering amplitudes at low energies cannot
be expressed as an expansion in the strong coupling constant because:
\begin{itemize}
\item At low energies the coupling constant becomes large.
\item We deal with bound states which require always to go beyond
a simple perturbation series.
\end{itemize}
So how can we calculate e.g. the cross-section of $\pi\pi$-scattering in QCD?

Two fundamental options are possible:
\begin{description}
\item[Lattice:]
This can be done using the methods developed by L\"uscher\cite{Luscher86}
via QCD calculations at finite volume. The energy spectrum at finite volume
depends on the $\pi\pi$ interaction. By measuring the 
lowest two-pion energy level,
the $\pi\pi$ $S-$wave scattering lengths $a_0^I$  can be obtained via
\be
E^I= 2 m_\pi-\frac{4\pi a_0^I}{m_\pi L^3}+\order (L^{-4})\,,
\ee
where $I$ denotes isospin. There are first lattice 
results available which make use of this relation \cite{scattlattice}.

\item[Effective Theory of QCD:]
Using the methods of {\em effective quantum field theory} (EFT) we can
also evaluate cross-sections in QCD at low energies. We replace
the QCD Lagrangian by the effective Lagrangian of Chiral Perturbation
Theory (CHPT). We replace the quarks and gluons by the degrees of freedom which are
relevant at low energies: $\pi$, $K$, $\eta$, $p$, $n$,\ldots.
This is the method most of this report is concerned with.
\end{description}

Chiral Perturbation Theory does reproduce for appropriately chosen coupling
constants the $S$-matrix elements of QCD\cite{weinberg79} and it allows
to make very sharp predictions in some cases. As an example, we quote 
 the recent result on the scattering length
combination $a_0^0-a_2^0$\cite{CGL},
\be
a_0^0-a_0^2 = 0.265\pm 0.004\,.
\ee
For earlier work at tree level and at one and two-loops accuracy, see 
\cite{pipi0,pipi1,pipi2}.

We will now present  the basics which  underly  the calculations 
of $\pi\pi\to\pi\pi$,
$\eta\to\pi\pi\pi$, etc. in this approach.

\subsection{Chiral Symmetry}

Consider QCD with 3 flavours, $u$, $d$ and $s$. Gluon interactions
do not change the helicity of a quark, only mass terms couple the left and
right handed helicity states. So in the limit where
$m_u=m_d=m_s=0$, the left-handed world cannot be turned into
the right handed world. Now if all the masses are zero, there is also
no way to distinguish the different flavours of quarks and we can continuously
rotate one into the other. The QCD Lagrangian is thus invariant under
\be
\left(\begin{array}{c}u\\d\\s\end{array}\right)_H
\quad\longrightarrow\quad
V_H\left(\begin{array}{c}u\\d\\s\end{array}\right)_H
\quad ;\quad H=L,R\,,
\ee
separately for both helicities, left (L) and right (R).
The matrices $V_H$ are general unitary $3\times3$ matrices. So we
have a global\footnote{The $U(1)$ parts of the groups play no role here,
but see the contributions by Bass\cite{Bass}, Shore\cite{Shore},
Michael\cite{Michael}, Kroll and Feldmann\cite{Kroll} for their effects.}
\be
SU(3)_L\times SU(3)_R
\ee
{\em chiral symmetry of QCD}\,. This symmetry has 16 conserved
currents and has thus 16 conserved charges. The vector charges
$Q_V^i = Q_L^i+Q_R^i$ also annihilate the
vacuum, but the axial charges $Q_A^i= -Q_L^i+Q_R^i$ produce a change on
the vacuum:
\be
Q_V^I |0\rangle = 0\quad\quad Q_A^I |0\rangle \ne 0\,.
\ee
The symmetry of the QCD Lagrangian is {\em not} the symmetry of
the vacuum. This is known as {\bf spontaneous breakdown of chiral
symmetry} or spontaneous chiral symmetry breaking ((S)$\chi$SB).

The 8 spontaneously broken axial symmetries require the existence of
8 Goldstone bosons which must be massless and pseudoscalar.
They are the pions, kaons and eta. The Goldstone bosons also have zero
interactions at zero energy. The expansion of CHPT is based on these facts.

Now, the fact that the quark masses are present and we can also have nonzero
energies allows nonzero values for masses and interactions.
E.g. the pion mass is an intricate mixture of the current quark mass,
{\em explicit $\chi$SB}, and the quark vacuum expectation value,
{\em spontaneous $\chi$SB}\cite{GMOR}.
\be
\label{GMOR}
m_\pi^2 = -(m_u+m_d)\langle 0 | \bar u u | 0\rangle/F^2\,.
\ee
Similarly the pion scattering lengths do not vanish\cite{pipi0}
\be
a_0^0 = \frac{7 m_\pi^2}{32\pi F_\pi^2} = 0.159\,.
\ee
These predictions are not exact, they are the first term in a series expansion
of momenta and the quark masses.

\subsection{A systematic expansion}

The amplitudes are expanded in powers of momenta and quark masses.
The price paid for the generality is that order by order we
get more terms,
\be
\lag_{\mbox{\small CHPT}} = \lag_2+\lag_4+\lag_6+\ldots\, ,
\ee
where $\lag_n$ contains terms with $n$-derivatives or equivalent.
In amplitudes the derivatives turn into momenta such that the expansion
converges at low momenta. As said earlier, quark masses are counted as
two powers of momentum, this is because Eq. (\ref{GMOR}) sets a quark
mass to the pion mass squared. The structure of the Lagrangians $\lag_n$ is
fixed by chiral symmetry but the coupling constants are not.

In the strong and semileptonic mesonic sector, these Lagrangians are known
and the parameters have conventional names:
\begin{description}
\item[$\lag_2$:] $F_0, B_0$ \cite{weinberg68}
\item[$\lag_4$:] $L_1,\ldots,L_{10}$ \cite{GL1}
\item[$\lag_6$:] $C_1,\ldots,C_{90}$  \cite{BCElag}
\item[$\vdots$]
\end{description}
These coupling constants (LEC's) are free in chiral perturbation theory,
 but are in principle calculable from QCD\footnote{For some attempts to
determine them from lattice QCD, see Ref.~\cite{Wittig}.}.

The Lagrangians $\lag_{2,4,6,\ldots}$ allow to calculate S-matrix elements
order by order in powers of momenta and quark masses. The main point
is that order by order, only a limited number of terms can contribute,
 and the form of the contribution is completely fixed
by chiral symmetry. So an amplitude can be expanded in powers
\be
A = A_2+A_4+A_6+\cdots\,,
\ee
with
\begin{description}
\item[$A_2$:] tree graphs with vertices from $\lag_2$. The form of the
vertices is fixed by chiral symmetry.
\item[$A_4$:] tree graphs with vertices from $\lag_2$ and one vertex from
$\lag_4$ or loop graphs with vertices from $\lag_2$.
\item[$\vdots$]
\end{description}
This procedure is known as {\em Chiral Perturbation Theory} (CHPT).

In loop integrals one integrates over all momenta, from eV to the Planck mass
and beyond, yet we always use the vertices from the effective Lagrangian
constructed to reproduce the amplitude at low energies. This is not a
contradiction, the high energy behaviour of the loop integrals
is irrelevant in the sense that this is the part that is absorbed
by the coupling constants. In this way, different regularizations
can be used to cut-off the integrals and the difference in different
ways of doing it is absorbed by using different values of the
coupling constants. The final result is always independent of the
regularization used. In practice we nearly always use
dimensional regularization
since it preserves chiral symmetry throughout the calculation reducing the need
for technically complicated subtractions to restore symmetries.

Many applications exist, see e.g. the miniproceedings of the Bad Honnef
meetings and the review articles\cite{reviews} as well
as the lectures \cite{chptlectures}.

\section{$\mathbf{\eta\to\pi\pi\pi}$ at order $\mathbf{p^4}$}
\label{p4eta}

\subsection{Kinematics and Isospin Relations}

The $\eta$ is an isospin singlet and a pseudo-scalar. The decay into two pions
is forbidden by CP. CP violation in $\eta$ decays is discussed
by C.~Jarlskog and E.~Shabalin\cite{Jarlskog} and J.~Ng\cite{Ng}
in these proceedings.
Three pions in an angular momentum 0 configuration cannot have isospin zero
as well, but isospin 1 is allowed. This decay thus has to proceed via
isospin breaking effects. Electromagnetism is known to
play a fairly minor role here as discussed below, except via the kinematical
effects due to the charged and neutral pion mass difference. This decay
thus goes primarily through the strong isospin breaking part of QCD
\be
{\lag}_{\mbox{I\hskip-1ex/}} =
- \frac{1}{2}(m_u-m_d)\left(\bar u u -\bar d d\right)\,.
\ee
This itself has isospin 1 and there is thus to lowest order in isospin breaking
a relation between $\eta\to\pi^+\pi^-\pi^0$ and $\eta\to\pi^0\pi^0\pi^0$.

\begin{figure}
\begin{center}
\includegraphics{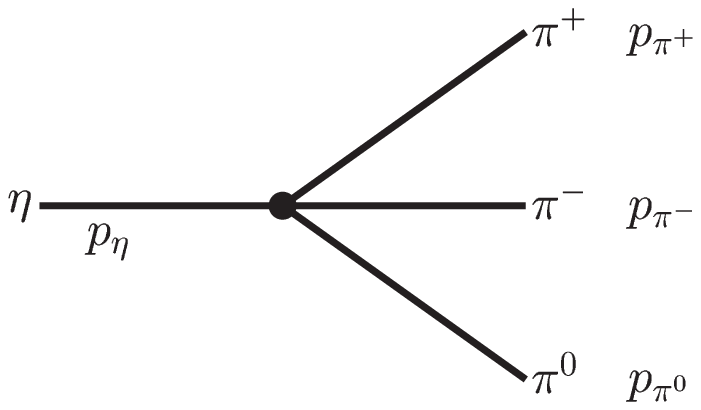}
\end{center}
\caption{\label{figkineta} The kinematical variables in the decay
$\eta\to\pi^+\pi^-\pi^0$.}
\end{figure}

The kinematics is depicted in Fig.~\ref{figkineta}.

The variables used
in the remainder are
\ba
s &=&\left(p_{\pi^+}+p_{\pi^-}\right)^2=\left(p_\eta-p_{\pi^0}\right)^2
\, ,\nonumber\\
t &=&\left(p_{\pi^-}+p_{\pi^0}\right)^2=\left(p_\eta-p_{\pi^+}\right)^2
\, ,\nonumber\\
u &=&\left(p_{\pi^+}+p_{\pi^0}\right)^2=\left(p_\eta-p_{\pi^-}\right)^2\,,
\ea
which satisfy
\be
s+t+u = m_\eta^2+2 m_{\pi^+}^2+m_{\pi^0}^2 \equiv 3 s_0\,.
\ee
The last equality is the definition of $s_0$.

The decay amplitude itself can be written as
\be
\langle \pi^0\pi^+\pi^-{\mbox {out}}|\eta\rangle = i\left(2\pi\right)^4 
\,\delta^4\left(p_\eta-p_{\pi^+}-p_{\pi^-}-p_{\pi^0}\right)
\,A(s,t,u)\,.
\ee
Charge conjugation requires this amplitude to be symmetric under the
interchange of $\pi^+$ and $\pi^-$ so we have
\be
A(s,t,u) = A(s,u,t)\,.
\ee
Using isospin as discussed above and labeling the three
momenta as $p_1,p_2,p_3$ and $s_i=\left(p_\eta-p_i\right)^2$, the amplitude
for the neutral decay
\be
\langle \pi^0\pi^0\pi^0{\mbox {out}}|\eta\rangle = i\left(2\pi\right)^4 
\,\delta^4\left(p_\eta-p_{1}-p_{2}-p_{3}\right)
\,\overline{A}(s_1,s_2,s_3)
\ee
satisfies
\be
\overline{A}(s_1,s_2,s_3) = A(s_1,s_2,s_3)+A(s_2,s_3,s_1)+A(s_3,s_1,s_2)\, ,
\ee
see e.g. \cite{Walker} for a detailed derivation of this result.
Experimentally, there are two main numbers to be kept in mind\cite{PDG00}.
The decay width from an overall fit
\be
\label{decayrate}
\Gamma(\eta\to\pi^+\pi^-\pi^0) = 271\pm25~eV\, ,
\ee
with a scale factor of $S=1.8$, and the ratio of neutral to charged decays
\be
r \equiv \frac{\Gamma(\eta\to\pi^0\pi^0\pi^0)}{\Gamma(\eta\to\pi^+\pi^-\pi^0)}
= 1.404\pm0.034\, ,
\ee
with a scale factor of $S=1.3$. The presence of these scale factors is an
indication of mutually incompatible experiments.

\subsection{Lowest order: ${p^2}$}

The lowest order CHPT contribution from the quark mass difference
was derived in the sixties using current algebra methods\cite{Cronin67}
and is
\be
\label{etaLO}
A(s,t,u) = \frac{\dsp B_0 (m_u-m_d)}{\dsp 3 \sqrt{3} F_\pi^2}
\left\{1+\frac{3(s-s_0)}{m_\eta^2-m_\pi^2}\right\} \,.
\ee
As announced, the decay amplitude is proportional to $m_u-m_d$. We can now
rewrite the amplitude in terms of $Q$, a combination of quark masses
which is experimentally more easily accessible,
\be
Q^2 \equiv \frac{m_s^2-\hat m^2}{m_d^2-m_u^2}\, ,
\ee
where $m_i$ stands for the up, down or strange current quark mass,
 and
\be
\hat m = \frac{1}{2} (m_u+m_d)\,.
\ee
The decay rate can then be written as
\be
\label{defM}
A(s,t,u) =\frac{1}{Q^2} \frac{m_K^2}{m_\pi^2}(m_\pi^2-m_K^2)\,
\frac{\dsp M(s,t,u)}{3\sqrt{3}F_\pi^2}\, ,
\ee
with at lowest order
\be
M(s,t,u) = \frac{\dsp 3 s-4m_\pi^2}{\dsp m_\eta^2-m_\pi^2}\,.
\ee
It can alternatively be rewritten in terms of
\be
R = \frac{m_s-\hat m}{m_d-m_u}
\ee
as
\be\label{eqR}
A(s,t,u) = -\frac{1}{4\sqrt{3}F_\pi^2 R}\left( 3 s- 4 m_\pi^2\right)\,.
\ee

So if we have a good theoretical description of $M(s,t,u)$, we get an accurate
determination of $Q$, since
\be
\Gamma\left(\eta\to3\pi\right) \propto
|A|^2\propto Q^{-4}\,.
\ee
Due to the high power, the error on $Q$ can be made much smaller than
its determination from meson mass ratios,
\be
Q^2 = \frac{m_K^2}{m_\pi^2}\frac{m_K^2-m_\pi^2}
{(m_{K^0}^2-m_{K^+}^2)_{{\tiny{\mbox{QCD}}}}}(1+O(m_{quark}^2)),
\ee
because this relation has  much larger electromagnetic 
corrections\cite{Dashenviolations}.

In order to show the accuracy which can be obtained, we can input two
different values of $Q$ or $R$.
We first use the value
\ba
R &=& 40.8\pm 3.2\quad\quad\mbox{\cite{Leutwyler96}}\, ,
\nonumber\\
\Gamma(\eta\to\pi^+\pi^-\pi^0) &\approx& 66~\mbox{eV}\, ,
\nonumber\\
r &=& 1.51\, . 
\ea
The second one is where we use instead the fit done at $p^6$ order
for the quark masses from the meson masses and the estimate
of the kaon electromagnetic mass difference including higher order
effects \cite{ABT},
\ba
Q &=& 20.0\pm1.5\quad \mbox{or}\quad R \approx 29\, ,
\nonumber\\
\Gamma(\eta\to\pi^+\pi^-\pi^0) &\approx& 140~\mbox{eV}\,.
\ea
It is clear that a good understanding of $M(s,t,u)$ can lead to a very
accurate determination of $Q$ and $R$.

\subsection{Electromagnetic effects}

The lowest order contribution does
not contribute directly to the decay rate. It only comes in via the
electromagnetic mass difference of pions. This is very relevant for
the actual values of the kinematical variables $s,t,u$ in terms
of the measured momenta, but this can be included in a simple
fashion. That this decay could not be explained by its electromagnetic
contributions has been known for a long time already\cite{dashen2}.
 
The calculation of the purely electromagnetic contribution to the decay has
also been pushed to one-loop level\cite{BaurKambor} and the total effect
remains small.

The decay
with an extra photon in the final state has been estimated as well.
This could have had potentially a large impact, since it can proceed
without extra isospin breaking. Fortunately, both a simplified analysis
using Low's theorem\cite{BramonGosdzinsky} and a full analysis at one
loop in CHPT\cite{Neufeldetal}
lead to a rather small result of the order of 1\% of
the decay width of $\eta\to3\pi$.

The correct incorporation of the electromagnetic corrections also
tells us which value for the pion mass should be introduced in the
expression for the amplitude. It is the same for both decays and is
the neutral pion mass but we need to include the correct physical
masses in the phase-space integrals. This follows from
adding to $\lag_2$ the effective electromagnetic term
\be
\lag_2\longrightarrow \lag_2+C\langle QUQU^\dagger\rangle\, ,
\ee
with
\be
\langle QUQU^\dagger\rangle = -\frac{2e^2}{F^2}\left(\pi^+\pi^-+K^+K^-\right)
+\frac{2e^2}{3F^2}\pi_3^2\pi^+\pi^-+\ldots \,.
\ee
There is no direct contribution to $\eta\rightarrow 3\pi$. The first term
leads to  an equal mass shift
for the charged pion and charged kaon in the chiral limit. This fact 
 is known as Dashen's theorem\cite{dashen}. The second term generates a 
contribution through mixing. This may appear to be negligibly small, 
because it is of order $e^2(m_d-m_u)$. However, this matters in 
the ratio $r$. Taking these corrections into account
tells us that the pion mass appearing in Eq. (\ref{eqR}) is
the neutral pion mass.

\subsection{Conclusions from the tree level}
\begin{itemize}
\item
The total rate is off by a factor of about 4, depending on the precise
inputs used.
\item
The ratio $r$ is near its experimental value. The reason for this
is discussed below.
\item
Second order isospin breaking effects are important in the phase-space.
Since the latter is small, small terms are important there - they
significantly lower $r$.
\item
The effective Lagrangian method is a very efficient tool to perform  
these calculations.
\end{itemize}

\subsection{One-loop contribution}

The amplitude can be expanded as
\be
A(s,t,u) = A_2(s,t,u)+A_4(s,t,u)+\ldots
\ee
$A_2$ has been calculated in the previous subsections. To evaluate
$A_4(s,t,u)$, one needs to evaluate one-loop diagrams with $\lag_2$
and tree diagrams with $\lag_4$. This was done in Ref.~\cite{GLeta}.
The decay rate can now be written in the form of Eq. (\ref{defM}).
 From the expression in Ref.~\cite{GLeta}, it can be seen that
the only free parameter coming in is $L_3$. This parameter can be determined
from $\pi\pi$-scattering\cite{GL1} or more accurately
from $K_{e4}$ decays\cite{ABT,ABT1}.

What one numerically finds is a very large enhancement over the lowest order
expression,
\be
\frac{\dsp \int dLIPS |A_2+A_4|^2}{\dsp \int dLIPS |A_2|^2} = 2.4\,,
\ee
 where $LIPS$ stands for Lorentz invariant phase-space.
This enhancement has as one major source the large $S$-wave final state
rescattering as was expected, see Ref.~\cite{Truong} and references therein.
In Sect. \ref{dispeta} we will come back to estimates of this contribution
to higher orders. This contributions is about half of the total enhancement
found in Ref.~\cite{GLeta}.

\subsection{$r$ and slope parameters}

\begin{figure}
\centerline{
\includegraphics[width=6cm]{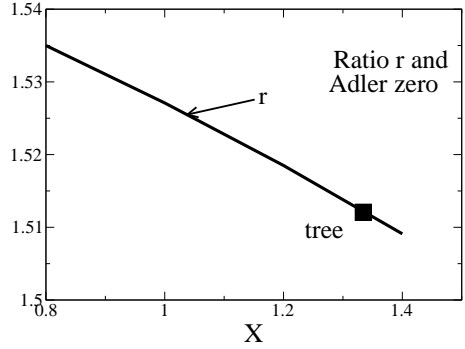}}
\caption{\label{figr}
The dependence of $r$ on $X$ as defined in Eq. (\ref{defXapprox}).}
\end{figure}

Can $r$ be very different from 1.5? The lowest order amplitude
has a zero for $s= (4/3) m_\pi^2$. Higher order corrections are not
expected to remove the zero and also not to move it very much. We can thus
expect that the amplitude will remain roughly of the form
\be
\label{defXapprox}
A(s,t,u)\approx b\left(s-X m_\pi^2\right)\,,
\ee
with $X$ about 4/3. Given this form  we see from Fig.~\ref{figr}
that unless the zero moves very much, the ratio $r$ will remain close
to 1.5.
The reason for this relation between $X$ and the ratio $r$ is
the isospin relation between the amplitudes for the neutral and
charged decays.

At $\order(p^4)$
the ratio $r$ is changed from the tree level prediction.
It is now lowered to 1.43, bringing it into nice agreement with
the observed ratio. Again, the relation between $X$, or more generally
the slope measurements in the Dalitz plot, and $r$  follows
from isospin. This is discussed in detail in Ref.~\cite{GLeta}.

\subsection{$Q$ at order $p^4$}

\begin{figure}
\begin{center}
\includegraphics[width=7cm ]{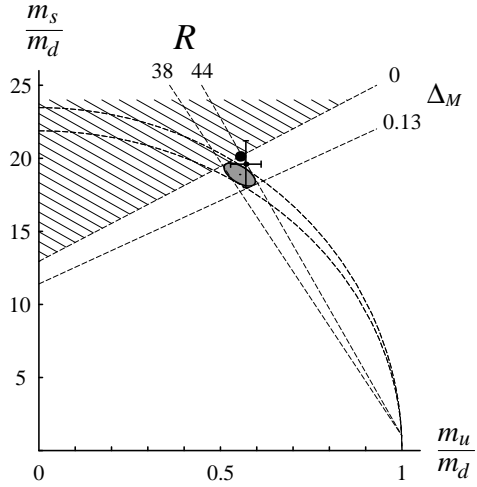}
\end{center}
\caption{\label{figQ96} Leutwyler's ellipse and the constraints
from mass differences at $\order(p^4)$. Taken from
Ref.~\cite{Leutwyler96}.}
\end{figure}

\begin{figure}
\begin{center}
\includegraphics[width=10cm ]{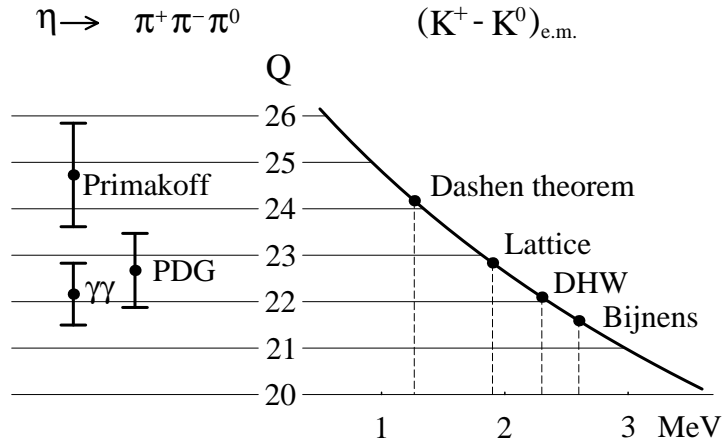}
\end{center}
\caption{\label{figQ96_2} The importance of $\eta\to3\pi$ for the
measurement of $Q$. Taken from
Ref.~\cite{Leutwyler96}.}
\end{figure}

The quark mass ratio $Q$ determines the major semi axis of Leutwyler's  ellipse
in the $m_s/m_d$ and $m_u/m_d$ plane\cite{Leutwyler96},
\be
\frac{m_u^2}{m_d^2}+\frac{1}{Q^2}\frac{m_s^2}{m_d^2} = 1\,.
\ee
The situation at order $p^4$ is shown in Fig.~\ref{figQ96}.
The role that a better understanding of $\eta\to3\pi$ has
in this context is shown in Fig.~\ref{figQ96_2},
where at order $p^4$ the measurement of $\eta\to3\pi$ is compared
with the determination from the $K^+$-$K^0$ mass difference. The different
points on the graph correspond to various estimates of the electromagnetic
part of the mass difference. Note that estimates of the $p^6$ contributions
\cite{ABT} indicate possibly even lower values of $Q$ than are shown here.

\section{$\mathbf{\eta\to\pi\pi\pi}$ beyond order $\mathbf{p^4}$}
\label{dispeta}

\subsection{The method and equations to solve}

We can now try to estimate even higher orders using dispersion relation
techniques. This should catch a large part of the higher order
corrections, because the large $\pi\pi$ rescattering part was
already a large part of the one-loop prediction. Unfortunately, for a general
three-body decay, this is not such an easy undertaking. There are
several subtleties involved here, as compared to the simple case of a two-body
final state, because one can have imaginary parts not only from
the scattering, but also from the unstable particle.
The latter difficulty can be avoided by studying
instead the scattering process $\pi\eta\to\pi\pi$ with an $\eta$ mass such that
the eta cannot decay. Then we only have $\pi\pi$ singularities in the
various channels to deal with. This procedure is how one can restrict oneself
to the influence of two-body scattering only. The general framework is
known as Khuri-Treiman equations \cite{KhuriTreiman} and was 
used in Ref.~\cite{KWW} to study higher order corrections to eta decay.

Here we will restrict ourselves to a simpler formalism that has been very
successful in the study of $\pi\pi$ scattering\cite{KnechtStern}.
The underlying observation is that there are no imaginary parts
from rescattering with angular momentum larger than or equal to
two up to $p^8$.
This together with the fact that $s+t+u$ is constant shows that to order $p^6$
the amplitude (\ref{defM}) can be written as
\be
\label{splitM}
M(s,t,u) = M_0(s)+(s-u)M_1(t)+(s-t)M_1(t)
+M_2(t)+M_2(u)-\frac{2}{3}M_2(s)\,.
\ee
The functions $M_I(s)$ correspond to isospin $I$ rescattering in the two
particles whose kinematics is described by $s$. The split into
these functions is not quite unique. There is some ambiguity
in the distribution of the polynomial terms over the various $M_I$,
 because $s+t+u$ is constant. The split (\ref{splitM}) is useful because:
\begin{itemize}
\item Dispersion problems are now reduced to a set of functions
of only {\em one} variable much simplifying the analysis.
\item This split is fully correct to two-loop order.
\item The main two-body rescattering corrections happen inside the
functions $M_I$.
\end{itemize}
This method was used to estimate higher order corrections to
$\eta\to3\pi$ in Ref.~\cite{AnisovichLeutwyler}.
A detailed description can be found in Ref.~\cite{Walker}.
With the approximations used in Ref.~\cite{KWW}, the two methods are in fact
entirely equivalent.

So we use here the form (\ref{defM}) and discuss the process
$\pi\eta\to\pi\pi$ with the final two pions in isospin $I$ for an eta mass
below the three pion threshold. This leads to the dispersion relations
\be
M_I(s) = \frac{1}{\pi}\int_{4m_\pi^2}^\infty\,ds^\prime\,
\frac{\im M_I(s^\prime)}{s^\prime-s-i\varepsilon}\, ,
\ee
up to subtractions, see below. To be more
precise, the integrand contains the discontinuity
\be
\im M_I(s^\prime)\longrightarrow \disc M_I(s) = \frac{1}{2i}\left(
M_I(s+i\varepsilon)-M_I(s-i\varepsilon)\right)\,.
\ee

We will need to subtract to be able to compare with the one-loop
expression. Checking the high-energy behaviour of the one-loop expressions
leads to three subtractions for $M_0$ and $M_2$, while two are
sufficient for $M_1$. So the equations we will need to solve
for $M_I$ are
\ba
M_0(s) &=& a_0+b_0 s+c_0 s^2+\frac{s^3}{\pi}\,\int\,\frac{ds^\prime}{s^{\prime3}}
\,\frac{\disc M_0(s^\prime)}{s^\prime-s-i\varepsilon}\,,
\nonumber\\
M_1(s) &=& a_1+b_1 s+\frac{s^2}{\pi}\,\int\,\frac{ds^\prime}{s^{\prime2}}
\,\frac{\disc M_1(s^\prime)}{s^\prime-s-i\varepsilon}\,,
\nonumber\\
M_2(s) &=& a_2+b_2 s+c_2 s^2+\frac{s^3}{\pi}\,\int\,\frac{ds^\prime}{s^{\prime3}}
\,\frac{\disc M_2(s^\prime)}{s^\prime-s-i\varepsilon}\,.
\ea

This leads to eight constants which should be determined. But the ambiguity in
the precise choice of the $M_I$ is visible here too. Up to the
same order in the $s,t,u$ expansion we have only four parameters
\be
M(s,t,u) = a + bs + c s^2 -d (s^2 + tu)\, ,
\ee
so there is some freedom in the choice. 

We can form some more convergent
combinations at one-loop as well. 
Comparing expressions one
sees that
\be
M_0(s)+\frac{4}{3}M_2(s)
\ee
and
\be
s M_1(s)+M_2(s) + s^2\frac{4 L_3-1/(64\pi^2)}{F_\pi^2(m_\eta^2-m_\pi^2)}
\ee
need only two subtractions. This allows to determine $c$ and
 $d$ in terms of $L_3$ and integrals over the discontinuities
via
\ba
c &=& c_0+\frac{4}{3} c_2 =
\frac{1}{\pi}\,\int\,\frac{ds^\prime}{s^{\prime3}}
\,\left\{\disc M_0(s^\prime)+\frac{4}{3}\disc M_2(s^\prime)\right\}
\,,
\nonumber\\
d &=& -  \frac{4 L_3-1/(64\pi^2)}{F_\pi^2(m_\eta^2-m_\pi^2)}
+\frac{1}{\pi}\,\int\,\frac{ds^\prime}{s^{\prime3}}
\,\left\{s^\prime\disc M_1(s^\prime)+\disc M_2(s^\prime)\right\}
\,.
\ea
So once we know $a$, $b$ and $L_3$, this forms a set of integral
equations that can be solved. Now we have to put in the knowledge of
the $\pi\pi$ phases. In a single channel problem without singularities
in the cross channel, this problem is simple and was solved by
Omn\`es long ago. Here we have cross-channel singularities
but they are again $\pi\pi$ scattering in an isospin 0,1,2 state.
The discontinuities thus obey\cite{Anisovich}
\be
\label{Mhat}
\disc M_I(s) = \sin \delta_I(s) e^{-i\delta_I(s)}
\left\{M_I(s)+\hat M_I(s)\right\}\,,
\ee
with $\hat M_I$ a consequence of the singularities in the $t$ and $u$
channel. They satisfy
\be
\label{Mhat2}
\hat M_I(s) = \sum_{n,I^\prime} \int_{-1}^{1}d\cos\theta\, \cos^n\theta
c_{nII^\prime} M_{I^\prime}(t)\, ,
\ee
where $t$ is the appropriate $t$ for a scattering of angle $\theta$ with
center of mass energy squared of $s$. The coefficients $c_{nII^\prime}$
can be found in \cite{Walker,Anisovich}.

So at least we have a principle solution. There are several
technical complications involved. After analytically continuing $m_\eta^2$
to its physical value, extra singularities appear, and we need $\hat M(s)$
for values of $s$ outside the physical domain, so singularities
in the relation of $t$ with $s$ and $\cos\theta$ appear. The integration
path needs to be chosen carefully to avoid these extra singularities.
Having solved this problem, one finds that the solution to the equations
is not unique. The problem is that homogeneous equations
of the type
\be
M_I(s) = \frac{1}{\pi}\int\,\frac{ds^\prime}{s^\prime-s-i\varepsilon}
\sin \delta_I(s) e^{-i\delta_I(s)}
\left\{M_I(s)+\hat M_I(s)\right\}\,
\ee
have nontrivial solutions - we need to pick out the
physically correct one. The solution is \cite{AnisovichLeutwyler}
to write a dispersion relation for a related function which only has
one solution. We remove the singularity in the direct channel
by dispersing instead 
\be
m_I(s) = \frac{M_I(s)}{\Omega_I(s)}\, ,
\ee
with
\be
\Omega_I(s) = \exp\left\{\frac{s}{\pi}\int \frac{ds^\prime}{s^\prime}
\,\frac{\delta_I(s^\prime)}{s^\prime-s-i\varepsilon}\right\}\,.
\ee
Similar arguments to above lead then to the dispersion relations
with a total of 4 subtraction constants for the $m_I$. They are, rewritten in
the $M_I$,
\ba
\label{finaldisp}
\frac{M_0(s)}{\Omega_0(s)} &=& \alpha_0+\beta_0 s+\gamma_0 s^2
+\frac{s^2}{\pi}\int\,ds^\prime
\frac{\sin\delta_0(s^\prime)\,\hat M_0(s^\prime)}
{|\Omega_0(s^\prime)| s^{\prime2} (s^\prime-s-i\varepsilon)}\,,
\nonumber\\
\frac{M_1(s)}{\Omega_1(s)} &=& \beta_1 s
+\frac{s}{\pi}\int\,ds^\prime
\frac{\sin\delta_1(s^\prime)\,\hat M_1(s^\prime)}
{|\Omega_1(s^\prime)| s^{\prime} (s^\prime-s-i\varepsilon)}\,,
\nonumber\\
\frac{M_2(s)}{\Omega_2(s)} &=& 
\frac{s^2}{\pi}\int\,ds^\prime
\frac{\sin\delta_2(s^\prime)\,\hat M_2(s^\prime)}
{|\Omega_2(s^\prime)| s^{\prime2} (s^\prime-s-i\varepsilon)}\,.
\ea

So basically we now have to find a parametrization of $\delta_{0,1,2}(s)$,
the $\pi\pi$ phaseshifts, and determine the subtraction constants.
After that the above equations can be solved numerically.
Two of the constants
can be determined similarly to $c$ and $d$ above, leading to
\ba
\gamma_0 &\approx& 0\,,
\nonumber\\
\beta_1 &\approx&  -  \frac{4 L_3-1/(64\pi^2)}{F_\pi^2(m_\eta^2-m_\pi^2)}\,.
\ea
The values of these two parameters compared with the one-loop expressions
are basically independent of the matching points. The other two
are discussed in the next section.

\subsection{Results}

The two papers \cite{AnisovichLeutwyler} and \cite{KWW} basically only differ
in the choice of the subtraction procedure. As explained above, the
formalism is quite different but the underlying approximations and assumptions
are the same.

Reference \cite{AnisovichLeutwyler} looks at the lowest order expression for
\be
M(s,t,u) = \frac{3s-4m_\pi^2}{m_\eta^2-m_\pi^2}\, , 
\ee
and notices that we can determine $\alpha_0$ and $\beta_0$ from
the place of the zero, the Adler zero happening at $s_A$, and the value
of the slope in $s$ at $s=s_A$.

\begin{figure}
\begin{center}
\includegraphics[width=7cm,height=6cm]{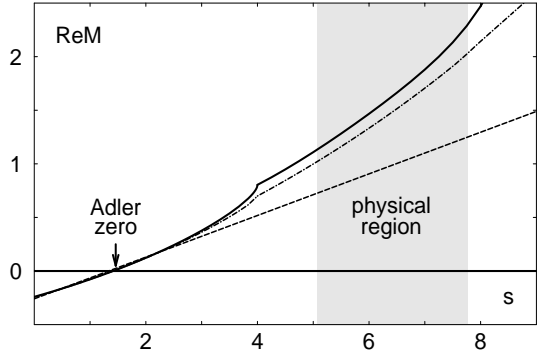}
\end{center}
\caption{\label{figMAW} The result of Ref.~\cite{AnisovichLeutwyler}
for $\re M(s,t,u)$ along the line $t=u$.}
\end{figure}

\begin{figure}
\begin{center}
\includegraphics[width=7cm,height=10cm]{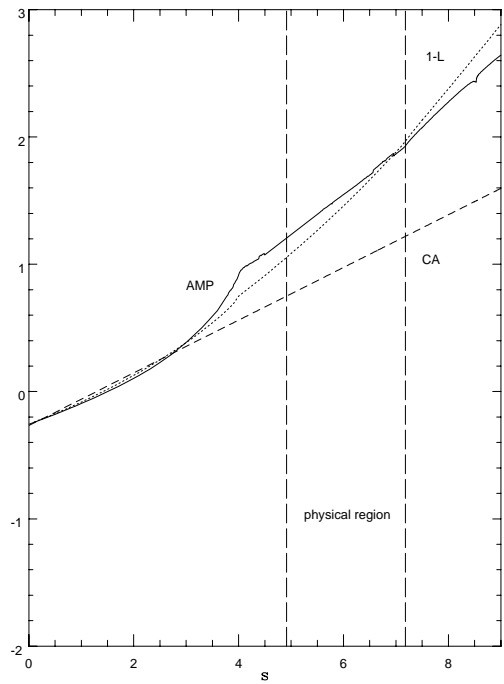}
\end{center}
\caption{\label{figMKWW} The result of Ref.~\cite{KWW}
for $\re M(s,t,u)$ along the line $t=u$.}
\end{figure}

Reference \cite{KWW} instead uses several values of $s,t,u$ to
fit the dispersive approximation to the one-loop result. The resulting
divergence in results is taken as an estimate of the theoretical error.
The final result for the enhancement over the one-loop result is quite
similar in both papers, but a closer look at the intermediate results
shows some discrepancies which need to be understood. The one thing which can
be compared are the published plots of $\re M(s,t,u)$ as a function
of $s$ along the line $t=u$.
These are shown in Fig.~\ref{figMAW} and \ref{figMKWW}.
Notice that while both have a significant enhancement, the slope is quite
different. The results of \cite{AnisovichLeutwyler} are always
above the one-loop result while the results of \cite{KWW} cross the
one-loop result at the end of the physical region. Given the similarity of
the methods and the inputs used, this type of discrepancies needs to be
understood, but it is obvious that a better experimental determination
of the slopes can distinguish between both calculations. It is therefore
very important that all slope parameters, see below for the definitions,
are well measured.

\begin{figure}
\begin{center}
\includegraphics[width=11cm]{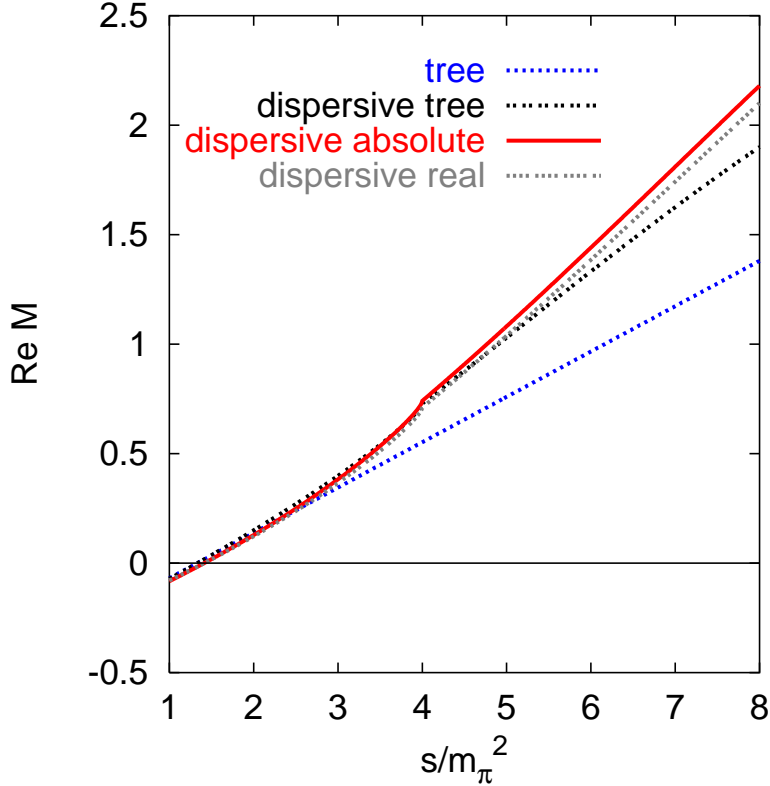}
\end{center}
\caption{\label{figmyeta} The value of $\re M(s,t,u)$ in a simplified
version of the dispersion relations with several different estimates
of the subtractions. $\alpha_0$ and $\beta_0$ are determined from
the tree level (dispersive tree)
the position of the one-loop Adler zero, 
and either
the absolute value of the slope (dispersive absolute) or its real part 
(dispersive real)
at one-loop. The tree level (tree) result is shown for comparison.}
\end{figure}

As a simple estimate of the uncertainties involved, we have used a simplified
version of the equations (\ref{finaldisp}) where we set the functions
$\hat M_I=0$. So we only consider rescattering effects in the direct channel.
We then fix the values of $\alpha_0$ and $\beta_0$, considered real, to
the tree level slope and Adler zero. In Fig.~\ref{figmyeta} the tree level
is shown (tree) and the dispersive
improved tree level (dispersive tree).
The two dispersive improvements of the one-loop result shown
are using 
the position of the one-loop Adler zero, 
and either
the absolute value of the slope (dispersive absolute) or its real part 
(dispersive real)
at one-loop to determine $\alpha_0$ and $\beta_0$.
It can be seen that these predictions lead to different
slopes in the physical regions, thus allowing the theoretical subtraction
procedure to be experimentally tested. Notice that we have shown only
one variation in the phase-space. The variation over all of phase-space
can of course be
used similarly.

The total enhancement found is fairly small, about 14\% in Ref.~\cite{KWW}
for the total rate
and 5\% in the center of the Dalitz plot in Ref.~\cite{AnisovichLeutwyler},
 with a total rate enhancement in very good agreement with Ref.~\cite{KWW}.

\subsection{Dalitz plot expansions or slope parameters}

\begin{figure}
\begin{center}
\includegraphics[width=9cm]{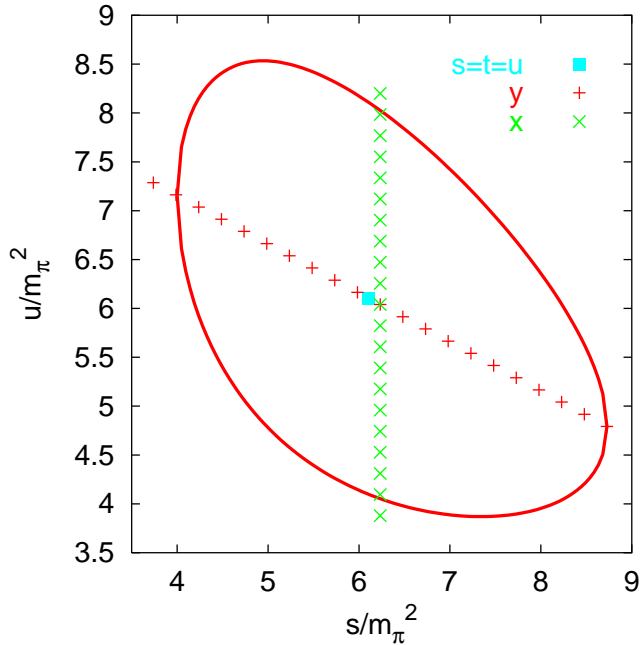}
\end{center}
\caption{\label{dalitzeta} The Dalitz plot or phase-space of
$\eta\to\pi^+\pi^-\pi^0$ decays. Shown are the boundaries of
the physical region, the variation with $x$ and $y$, see text,
and the point $s=t=u$.}
\end{figure}

Once the amplitude is known,
one can easily expand the Dalitz plot distributions
around the center of the Dalitz plot. Notice that some care is needed
to precisely define the center of the Dalitz plot, given the second
order isospin breakings discussed earlier. The point $s=t=u=s_0$
does not quite coincide with the point where all kinetic energies are equal
for the charged decay.
The usual definition of the slope parameters is in terms of the
kinematical variables $x$ and $y$,
\ba
x &=& \sqrt{3}\,\frac{T_+-T_-}{Q_\eta} = 
\frac{\sqrt{3}}{2 M_\eta Q_\eta} (u-t)\,,
\nonumber\\
y &=& \frac{3 T_0}{Q_\eta}-1\,
= \frac{3}{2 m_\eta Q_\eta}\left\{\left(m_\eta-m_{\pi^0}\right)^2-s\right\}-1
\, ,\nonumber\\
Q_\eta &=& m_\eta - 2 m_\pi^+ - m_\pi^0
\,.
\ea
$T_i$ stands for the kinetic energy of the pion of charge $i$.
In Fig.~\ref{dalitzeta} we have shown how $s$ and $u$ depend on
$x$ and $y$. The crosses correspond to $x=0$ and $y$ equal to
$-1$, $-0.9$,\ldots,$1$. The $\times$ correspond to $y=0$
and $x=-1,0.9,\ldots,1$.
The filled square indicates the
place where $s=t=u$. The difference with the point $x=y=0$
is again an indication of the
isospin breaking effects in the masses discussed earlier.

The measured distribution is now parametrized as follows. For the charged
decay, one takes
\be
\label{defdalitz}
1+ ay + by^2+c x^2\, ,
\ee
normalized to the center $x=y=0$ of the Dalitz plot, and
\be
1+g(x^2+y^2)
\ee
for the neutral decay. It is hard to compare with the experimental
results in the review of particle properties \cite{PDG00}.
They simply state that the assumptions made in the different measurements
are not compatible. The problem is that making assumptions on
the values of the quadratic slopes significantly alters the fitted result
of the others. In table \ref{tabtheory} we show the theoretical results
of tree level, one-loop, the dispersive results from \cite{KWW}
and the results of the tree dispersive and absolute dispersive
simplified estimates discussed above. Note that the one-loop
predictions of the neutral slope $g$ is not zero, contrary to the
statement made in
Refs. \cite{CB2,Amsler3}. 
The loops themselves do give a contribution, whereas the contributions
from the tree  and tadpole diagrams at $\order(p^4)$ happen to vanish.
The precise value is dependent on the procedure used and
is sensitive to fairly small changes. The number quoted corresponds to
the precise way the expressions are given in \cite{GLeta}.
We plan to come back to this issue in a later publication.

\begin{table}
\begin{center}
\begin{tabular}{|c|cccc|}
\hline
                      & $a$     & $b$  & $c$  & $g$\\
\hline
tree                  & $-1.00$ & 0.25 & 0.00 & $0.000$\\
one-loop \cite{GLeta} & $-1.33$ & 0.42 & 0.08 & 0.03$^a$\\
dispersive \cite{KWW} & $-1.16$ & 0.26 & 0.10 & $-0.014$ --- $-0.028$\\
tree dispersive       & $-1.10$ & 0.31 & 0.001& $-0.013$\\
absolute dispersive   & $-1.21$ & 0.33 & 0.04 & $-0.014$\\
\hline
\end{tabular}
\end{center}
\caption{\label{tabtheory} Theoretical results for the
slope parameters of the various approximations discussed in the text.
The spread gives an indication of the uncertainty, for $g$ it is about a factor
two, even if the central values agree much better, see e.g. the discussion
in \cite{KWW}. $^a$Ref.~\cite{GLeta} does not quote a numerical 
value for $g$. We have obtained $g=0.03$ from the representation 
of the amplitude provided  in \cite{GLeta}.}
\end{table}

Measurements as mentioned earlier are difficult to compare, but
several new results have been obtained recently. In particular, the
new Crystal Ball value for $g$ was announced at this
 meeting \cite{Nefkens}. Note that
the experiments usually quote $\alpha = g/2$.
For completeness the known experimental results
are given in table \ref{tabexperiment} for the charged decay
and table \ref{tabexperiment2} for the neutral decay.
\begin{table}
\begin{center}
\begin{tabular}{|c|ccc|}
\hline
                      & $a$     & $b$  & $c$ \\
\hline
Layter\cite{Layter}   & $-1.08\pm0.14$ & $0.034\pm0.027$&$0.046\pm0.031$  \\
Gormley\cite{Gormley} & $-1.17\pm0.02$ & $0.21\pm0.03$ & $0.06\pm0.04$ \\
Crystal Barrel\cite{Amsler} 
                      & $-0.94\pm0.15$ & $0.11\pm0.27$ &  \\
Crystal Barrel\cite{Amsler2}        
             & $-1.22\pm0.07$ & $0.22\pm0.11$ & 0.06 fixed \\
\hline
\end{tabular}
\end{center}
\caption{\label{tabexperiment} Experimental results for the
slope parameters in the charged decay.}
\end{table}
\begin{table}
\begin{center}
\begin{tabular}{|c|c|}
\hline
                      &  $g$\\
\hline
Alde\cite{Alde}        &  $-0.044\pm0.046$\\
Crystal Barrel\cite{Amsler3}  & $-0.104\pm0.039$\\
Crystal Ball\cite{Nefkens,CB2} & $-0.062\pm0.008$\\
\hline
\end{tabular}
\end{center}
\caption{\label{tabexperiment2} Experimental results for the
slope parameter in the neutral decay. We have added in quadrature the
statistical and systematical errors quoted in \cite{Amsler3}.}
\end{table}

Notice that while the overall agreement is all right, there is a problem
in obtaining a value of $g$ which is compatible with the newer experiments.
The theoretical evaluation of $g$ suffers from large cancellations.

\section{Dalitz and double Dalitz decays}
\label{dalitzdecay}
\label{doubledalitz}

\begin{figure}
\begin{center}
\includegraphics{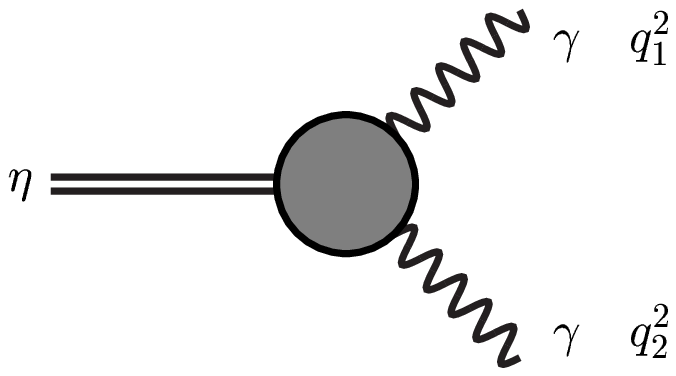}
\end{center}
\caption{\label{figetaoffshell} The $\eta\gamma\gamma$ vertex with
photons of mass squared $q_1^2$ and $q_2^2$.} 
\end{figure}

This section is a shortened version of the work
presented  in Ref.~\cite{BijnensPersson}. The underlying question is
how well do we understand the couplings of the $\eta$ to two photons.
In table \ref{tab:gg} we show the (expected) branching ratios
for all of the decays with two photons. The decays with one lepton-antilepton
pair are known as Dalitz decays, those with two pairs as double Dalitz decays.
\begin{table}
\begin{center}
\begin{tabular}{|l|c|}
\hline
Decay & Branching ratio\\
\hline
$\eta\to\gamma\gamma$          & $39.33\pm0.25~\%$\\
$\eta\to\gamma e^+e^-$         & $(4.9\pm1.1)\cdot10^{-3}$\\
$\eta\to\gamma \mu^+\mu^-$     & $(3.1\pm0.4)\cdot10^{-4}$\\
$\eta\to e^+ e^- e^+e^-$       & $\sim6\cdot10^{-5}$\\
$\eta\to e^+e^- \mu^+\mu^-$    & $\sim2\cdot10^{-6}$\\
$\eta\to \mu^+\mu^- \mu^+\mu^-$& $\sim10^{-8}$\\
\hline
\end{tabular}
\end{center}
\caption{\label{tab:gg} The measured or expected branching ratios
for various decays of $\eta$ to two (possibly off-shell)
 photons.}
\end{table}
These decays allow us to study the {\em off-shell} structure of the
$\eta\gamma\gamma$ vertex as shown in Fig.~\ref{figetaoffshell}.
The question is: how does this form-factor $F(q_1^2,q_2^2)$ 
behave over the entire range
of off-shell masses for the photons? Is it of order 1, w.r.t. to on-shell
photons, can it be described by adding vector meson propagators
in the photon legs? 
One would also like to know the  approach to
the perturbative
QCD result (see e.g. Ref.~\cite{formasympt})
\be
\label{shortQCD}
 F(-Q^2,-Q^2)=
\frac{8\pi^2 F_\pi^2}{3 Q^2} + \ldots\, ,Q^2\rightarrow \infty\, .
\ee
This type of form-factors are one of the places where experiments are
possible for the same underlying physical process over the entire
regime, ranging from the fully perturbative to the fully nonperturbative
one. The large and intermediate values of $|q_1^2|$ and $|q_2^2|$
can be studied in tagged photon-photon collisions while
the small timelike values can be studied in $\eta$-decays.

To study how well $\eta$-decays can see variations in the form-factor,
 in Ref.~\cite{BijnensPersson} four trial form-factors were investigated,
\ba
\label{form1}
F(q_1^2,q_2^2) &=& 1\, ,\\
\label{form2}
&=& \frac{m_V^4}{\left(m_V^2-q_1^2\right)\left(m_V^2-q_2^2\right)}\, ,\\
\label{form3}
&=& \frac{m_V^2}{\left(m_V^2-q_1^2-q_2^2\right)}\, ,\\
\label{form4}
&=& \frac{m_V^4-\left(4\pi^2F_\pi^2/3\right)\left(q_1^2+q_2^2\right)}
{\left(m_V^2-q_1^2\right)\left(m_V^2-q_2^2\right)}\, .
\ea
The first form is included to check how well a deviation
from a pointlike eta can actually be measured. The second one
is the standard double vector meson dominance model, while the third
was a variation that reproduced single vector meson dominance, while
numerically approaching Eq. (\ref{shortQCD}) quite well.
The last form was suggested in Ref.~\cite{form4try}, see also
 Ref.~\cite{form4extra}, Knecht and Nyffeler in Ref.~\cite{g-2} 
as well as the contribution by Feldmann and Kroll 
to this conference\cite{Kroll}.
One important reason why we like to know this form-factor better is that
pseudoscalar exchange is the major contribution to the
hadronic light-by-light scattering part of the muon anomalous magnetic
moment as studied in Ref.~\cite{g-2}.
The contribution of the exchange of $\pi^0$, $\eta$ and $\eta^\prime$
using the
different form-factors to $a_\mu = (g_\mu-2)/2$
is \cite{BijnensPersson,g-2} $7.9\cdot10^{-10}$ for form-factor (\ref{form2}),
$9.7\cdot10^{-10}$ for (\ref{form3}) and
$10.9\cdot10^{-10}$ for (\ref{form4}). It is thus clearly important for future
measurements of $a_\mu$ that this form-factor becomes experimentally
more constrained.

In tables \ref{tab:dalitz} to \ref{tab:doubledalitz2}
we show the results \cite{BijnensPersson} for the various decays 
as a function of the
electron-positron mass lower bound as well the muonic cases for all of
phase-space.
\begin{table}
\begin{center}
\begin{tabular}{|c|c|ccc|}
\hline
Decay & $m_{e^+e^-}|_{\mbox{min}}$ &
 (\ref{form1})  & (\ref{form2}),(\ref{form3}) & (\ref{form4}) \\
\hline
$\eta\to\gamma e^+e^-$ & 
  $2 m_e$&$1.62\cdot10^{-2}$& $1.67\cdot10^{-2}$& $1.66\cdot10^{-2}$\\
& 50 MeV&$4.6\cdot10^{-3}$& $5.1\cdot10^{-3}$& $5.0\cdot10^{-3}$\\
& 200 MeV&$8.6\cdot10^{-4}$& $11.5\cdot10^{-4}$& $10.9\cdot10^{-4}$\\
& 300 MeV&$2.2\cdot10^{-4}$& $3.6\cdot10^{-4}$& $3.3\cdot10^{-4}$\\
& 400 MeV&$3.0\cdot10^{-5}$& $6.3\cdot10^{-5}$& $5.6\cdot10^{-4}$\\
\hline
$\eta\to\gamma \mu^+\mu^-$ 
& --- &$5.5\cdot10^{-4}$& $7.7\cdot10^{-4}$& $7.3\cdot10^{-4}$\\
\hline
\end{tabular}
\end{center}
\caption{\label{tab:dalitz} The branching ratios for the single Dalitz decays
relative to $\eta\to\gamma\gamma$
as a function of the lower limit of the lepton pair mass for the different
form-factors defined in the text.}
\end{table}
\begin{table}
\begin{center}
\begin{tabular}{|c|c|ccc|}
\hline
Decay & $m_{e^+e^-}|_{\mbox{min}}$ &
 (\ref{form2})  & (\ref{form3}) & (\ref{form4}) \\
\hline
$\eta\to e^+e^-\mu^+\mu^-$ & 
  $2 m_e$&$5.6\cdot10^{-6}$& $5.6\cdot10^{-6}$& $5.3\cdot10^{-6}$\\
& 50 MeV&$11.6\cdot10^{-7}$& $11.7\cdot10^{-7}$& $10.9\cdot10^{-7}$\\
& 200 MeV&$5.0\cdot10^{-8}$& $5.2\cdot10^{-8}$& $4.7\cdot10^{-8}$\\
& 300 MeV&$2.8\cdot10^{-10}$& $2.9\cdot10^{-10}$& $2.6\cdot10^{-10}$\\
\hline
\end{tabular}
\end{center}
\caption{\label{tab:doubledalitz} The branching ratios for the
double Dalitz decay $\eta\to e^+e^-\mu^+\mu^-$
relative to $\eta\to\gamma\gamma$
as a function of the lower limit of the electron positron
pair mass for the different
form-factors defined in the text.}
\end{table}
\begin{table}
\begin{center}
\begin{tabular}{|c|c|ccc|}
\hline
Decay & $m_{e^+e^-}|_{\mbox{min}}$ &
 (\ref{form2})  & (\ref{form3}) & (\ref{form4}) \\
\hline
$\eta\to e^+e^-e^+e^-$ & 
  $2 m_e$&$6.7\cdot10^{-5}$& $6.7\cdot10^{-5}$& $6.6\cdot10^{-5}$\\
& 50 MeV&$4.4\cdot10^{-6}$ & $4.4\cdot10^{-6}$& $4.3\cdot10^{-6}$\\
& 100 MeV&$7.4\cdot10^{-7}$& $7.5\cdot10^{-7}$& $7.2\cdot10^{-7}$\\
& 200 MeV&$2.7\cdot10^{-9}$& $2.8\cdot10^{-9}$& $2.6\cdot10^{-9}$\\
\hline
$\eta\to \mu^+\mu^-\mu^+\mu^-$ & 
  --- &$9.2\cdot10^{-9}$& $9.4\cdot10^{-9}$& $8.5\cdot10^{-9}$\\
\hline
\end{tabular}
\end{center}
\caption{\label{tab:doubledalitz2} The branching ratios for the
double Dalitz decays $\eta\to \mu^+\mu^-\mu^+\mu^-$ and $\eta\to e^+e^-e^+e^-$
relative to $\eta\to\gamma\gamma$
as a function of the lower limit of the electron positron
pair mass for the different
form-factors defined in the text. The last line was not published earlier.}
\end{table}
It turns out that it is surprisingly difficult to see the difference
between the various models of the form-factors in $\eta$ decays, but
some constraints are possible. The form-factors which
are different in single Dalitz decays should be easily distinguishable,
 but more subtle differences, as the one between (\ref{form2})
and (\ref{form3}) - which only become visible in double Dalitz decays -
 will be difficult to see. Given the importance of these results, the
measurements should be performed. The differences between the effects
of the the form-factors in (\ref{form1}) to (\ref{form4})
is  $\order(p^6)$. Form-factors  (\ref{form2}) (\ref{form3}) only differ
at  $\order(p^8)$. The differences are a good indication of the importance
of the various orders.

\section{$\mathbf{\eta\to\pi^0\gamma\gamma}$}
\label{etapigg}

This decay is only touched upon in this report. A more comprehensive discussion
can be found in the review by Ametller\cite{Ametller} and in the references.
The main theory paper underlying this decay is \cite{ABBC}, and the main
interest is that this decay is a window on rather high order corrections
in CHPT. The $\order(p^4)$ result is small for several reasons.
The amplitude has two possible Lorentz structures. In the
gauge $\epsilon_i\cdot q_j=0$, with $\epsilon_i$ and $q_i$ the polarization
vector and momentum of photon $i$, the amplitude can be written
as
\be
M(\eta\to\pi^0\gamma\gamma) =
A q_1\cdot q_2\,\epsilon_1\cdot\epsilon_2
- B\left(\epsilon_1\cdot\epsilon_2\, P_\eta\cdot q_1\, P_\eta\cdot q_2
        +P_\eta\cdot\epsilon_1\,P_\eta\cdot\epsilon_2\,q_1\cdot q_2\right)\,.
\ee
At \order$(p^4)$ only the $A$ part of the amplitude is nonzero
and it is small. The pion loop is suppressed by isospin breaking
and the kaon loop is suppressed by $4m_K^2$ and an extra factor of
$1/12$ in the integral. The contributions are
\ba
\label{p4etapgg}
\pi\mbox{-loops} & & 0.84\cdot10^{-3}~\mbox{eV}\,,
\nonumber\\
K\mbox{-loops} & & 2.45\cdot10^{-3}~\mbox{eV}\,,
\nonumber\\
\mbox{sum} & & 3.9\cdot10^{-3}~\mbox{eV}\,,
\ea
rather far from the old experimental value of $0.84\pm0.18$~eV\cite{PDG00}.
Newer experimental limits come from
Novosibirsk\cite{epggexp} and the Crystal Ball\cite{Nefkens}.
The latter gives $0.42\pm0.14$~eV. The result of Eq. (\ref{p4etapgg})
is therefore more than two orders of magnitudes below the measured width.

\begin{figure}
\begin{center}
\includegraphics{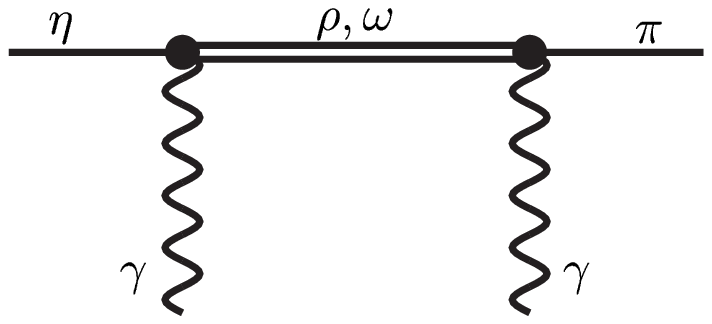}
\end{center}
\caption{\label{figepgg} The vector  meson exchange contribution
to $\eta\to\pi^0\gamma\gamma$.}
\end{figure}

We can also add vector meson exchange as depicted in Fig.~\ref{figepgg}.
This starts at $\order(p^6)$ and has contributions to
all orders. Restricting the exchange of $\omega$ and
$\rho$ to $p^6$ terms only leads to
$\Gamma_{\rho,\omega}^{(6)} = 0.18$~eV and if all orders are kept
\be
\Gamma_{\mbox{VMD}} = 0.31~\mbox{eV}\,,
\ee
much larger than the suppressed $p^4$ contributions. In fact, there is
one more contribution with well determined signs which appears at
$\order(p^8)$. Here there is no isospin suppression of the pion-loops any
longer nor any factors like the $1/12$ at $\order(p^4)$. As a result
the $p^8$ contribution from the double WZW vertex one-loop diagram
is as large as the $p^4$ one. Given the structure of the amplitude,
interference effects are quite strong so that if we put
in the loops plus the VMD estimates we obtain the estimate\cite{ABBC}
\be
\Gamma \approx 0.42~\mbox{eV}\,.
\ee
We can now add the contribution from $a_0$ and $a_2$ exchange,
the absolute values are experimentally known but the signs are free.
This leads to
\be
\Gamma \approx 0.42\pm0.20~\mbox{eV}\,.
\ee
in very nice agreement with the preliminary Crystal Ball result.
Studies of distributions in the decays will allow to disentangle
various contributions to this decay.
It should be noted that studies of this decay mode have
also been performed in the ENJL model\cite{epggenjl}
and by adding other resonances than the ones referred to above\cite{Koetal}.
There also exist some recent quark model studies\cite{quarkmodel}.

\section{Conclusions}
\label{conclusions}

We have discussed several eta decays from the point of view of Chiral
Perturbation Theory and various possible enhancements. The main conclusion
is that more and precise measurements are very much needed.

$\underline{\eta\to3\pi}:$ This decay has the potential to deliver
accurate quark mass ratios. For this we need better measurements
of the slope parameters in the Dalitz plot and a better determination of
the ratio $r$ between the neutral and charged decay mode. These will
allow to put the theoretical calculations of the decay rate on the
experimentally verified footing needed to obtain accurate quark mass ratios.
A possible discrepancy with the slope $g$ in the neutral decay should be
checked theoretically and further experimental confirmation
would be welcome.

$\underline{\eta\to\gamma^*\gamma^*}:$ The double off-shell form-factors
are needed in these double Dalitz decays
and provide a useful insight into short-distance long-distance transitions
in QCD. They are also needed input in precision calculations for the
muon anomalous magnetic moment. It turned out to be surprisingly
difficult to see these differences in $\eta$-decays, but it is an
important measurement to be performed.

$\underline{\eta\to\pi^0\gamma\gamma}:$ The existing discrepancy with theory
seems to be on the way to be resolved but we need confirmation of this and
the study of distributions will allow different contributions to this
decay mode to be distinguished. It is a decay mode where the usually
dominant contributions CHPT at tree level and one-loop are very small.
It thus provides a rare window on higher order contributions.

\section*{Acknowledgments}
This work is supported by the Swedish Research Council,
the Swiss National Science Foundation and
the European Union TMR network, Contract No. ERB\-FMRX--CT980169 
(EURODAPHNE), and BBW-Contract No. 97.0131. We thank 
the local organizing committee for
a well run and stimulating meeting and the sponsors,
the Swedish Research Council, the Royal Swedish Academy 
of Sciences through its Nobel Institute
for Physics and the Swedish Foundation for International Cooperation
in Research and Higher Education. 
We thank Heiri Leutwyler for reading the manuscript.


\begin{thebibliography}{99}

\bibitem{Holstein} Holstein, B., these proceedings [hep-ph/0112150]

\bibitem{Ametller} Ametller, Ll., these proceedings [hep-ph/0111278]

\bibitem{Luscher86}
Luscher, M.,
Commun.\ Math.\ Phys.\  {\bf 105}, 153 (1986).

\bibitem{scattlattice}
Aoki, S. {\it et al.}  [CP-PACS Collaboration],
Nucl.\ Phys.\ Proc.\ Suppl.\  {\bf 106} (2002) 230
[arXiv:hep-lat/0110151];\\
Liu, C. a., Zhang, J. h., Chen, Y. and Ma, J. P.,
[hep-lat/0109020];\\
Liu, C. a., Zhang, J. h., Chen, Y. and Ma, J. P.,
[hep-lat/0109010].

\bibitem{weinberg79}
Weinberg, S.,
Physica A {\bf 96}, 327 (1979);\\
Leutwyler, H.,
Annals Phys.\  {\bf 235}, 165 (1994)
[hep-ph/9311274].

\bibitem{CGL}
Colangelo, G., Gasser, J. and Leutwyler, H.,
Phys.\ Lett.\ B {\bf 488}, 261(2000)
[hep-ph/0007112].

\bibitem{pipi0}
Weinberg, S.,
Phys.\ Rev.\ Lett.\  {\bf 17}, 616 (1966).

\bibitem{pipi1}
Gasser, J. and Leutwyler, H.,
Phys.\ Lett.\ B {\bf 125}, 325 (1983).

\bibitem{pipi2}
Bijnens, J., Colangelo, G., Ecker, G., Gasser, J. and Sainio, M. E.,
Phys.\ Lett.\ B {\bf 374}, 210 (1996)
[hep-ph/9511397];
Nucl.\ Phys.\ B {\bf 508}, 263 (1997)
[Erratum-ibid.\ B {\bf 517}, 639 (1997)]
[hep-ph/9707291].

\bibitem{Bass} Bass, S. D., these proceedings, [hep-ph/0111180].

\bibitem{Shore} Shore, G., these proceedings [hep-ph/0111165].

\bibitem{Michael} Michael, C., these proceedings [hep-lat/0111056].

\bibitem{Kroll}
Feldmann, T. and Kroll, P.,
these proceedings,
[hep-ph/0201044].

\bibitem{GMOR}
Gell-Mann, M., Oakes, R. J. and Renner, B.,
Phys.\ Rev.\  {\bf 175}, 2195 (1968).

\bibitem{weinberg68}
Weinberg, S.,
Phys.\ Rev.\  {\bf 166}, 1568 (1968).

\bibitem{GL1}
Gasser, G. and Leutwyler, H.,
Nucl.\ Phys.\ B {\bf 250}, 465 (1985).

\bibitem{BCElag}
Bijnens, J., Colangelo, G. and Ecker, G.,
Annals Phys.\  {\bf 280} 100 (2000).
[hep-ph/9907333].

\bibitem{Wittig}
Heitger, J., Sommer, R. and Wittig, H.  [ALPHA Collaboration],
Nucl.\ Phys.\ B {\bf 588} 377 (2000)
[hep-lat/0006026];\\
Irving, A. C., McNeile, C., Michael, C., Sharkey, K. J. and Wittig, H.  [UKQCD
                  Collaboration],
Phys.\ Lett.\ B {\bf 518}, 243 (2001)
[hep-lat/0107023].


\bibitem{reviews}
Bijnens, J. and Mei\ss ner, U.-G., Workshop on the Standard Model
at low Energies, ECT$^*$, Trento, 1996, Miniproceedings [hep-ph/9606301];\\
Bijnens, J. and Mei\ss ner, U.-G., Workshop on
Chiral Effective Theories,
Bad Honnef, 1998, Miniproceedings [hep-ph/9901381];\\
Bijnens, J., Meissner, U.-G. and Wirzba, A.,
Workshop on
Effective Field Theories of QCD, Bad Honnef, 2001,
[hep-ph/0201266];\\
Ecker, G.,
Prog.\ Part.\ Nucl.\ Phys.\  {\bf 35}, 1 (1995)
[hep-ph/9501357];\\
Bernard, V., Kaiser, N. and Meissner, U.-G.,
Int.\ J.\ Mod.\ Phys.\ E {\bf 4}, 193 (1995)
[hep-ph/9501384].

\bibitem{chptlectures}
Pich, A., Lectures at Les Houches Summer School in
Theoretical Physics, Session 68: Probing the Standard Model of Particle
Interactions, Les Houches, France, 28 Jul - 5 Sep 1997,
[hep-ph/9806303];\\
Ecker, G.,
Lectures given at Advanced School on Quantum Chromodynamics (QCD 2000),
Benasque, Huesca, Spain, 3-6 Jul 2000,
[hep-ph/0011026].

\bibitem{Jarlskog} Jarlskog, C. and Shabalin, E., these proceedings.

\bibitem{Ng} Ng, J. N., these proceedings.

\bibitem{Walker}
Walker, M., ``$\eta\to3\pi$'', Master Thesis, Bern University (1998).
Can be obtained from {\tt http://www-itp.unibe.ch/research.shtml\#diploma}.

\bibitem{PDG00}
Groom, D. E. {\it et al.}  [Particle Data Group Collaboration],
Eur.\ Phys.\ J.\ C {\bf 15}, 1 (2000).

\bibitem{Cronin67} 
Bell, J. S. and Sutherland, D. G.,
Nucl.\ Phys.\ B {\bf 4}, 315 (1968);\\
Cronin, J. A.,
Phys.\ Rev.\  {\bf 161}, 1483 (1967).

\bibitem{Dashenviolations} 
Donoghue, J. F., Holstein, B. R. and Wyler, D.,
Phys.\ Rev.\ D {\bf 47}, 2089 (1993);\\
Bijnens, J.,
Phys.\ Lett.\ B {\bf 306}, 343 (1993)
[hep-ph/9302217];\\
Bijnens, J. and Prades, J.,
Nucl.\ Phys.\ B {\bf 490}, 239 (1997)
[hep-ph/9610360].

\bibitem{Leutwyler96}
H.~Leutwyler, H.,
Phys.\ Lett.\ B {\bf 378}, 313 (1996)
[hep-ph/9602366].

\bibitem{ABT}
Amoros, G., Bijnens, J. and P.~Talavera, P.,
Nucl.\ Phys.\ B {\bf 602}, 87 (2001)
[hep-ph/0101127].

\bibitem{dashen2}
Sutherland, D. G.,
Nucl.\ Phys.\ B {\bf 2},  433(1967);\\
Dittner, P., Eliezer, S. and Dondi, P. H.,
Phys.\ Rev.\ D {\bf 8}, 2253 (1973).

\bibitem{BaurKambor}
Baur, R., Kambor, J. and Wyler, D.,
Nucl.\ Phys.\ B {\bf 460}, 127 (1996).
[hep-ph/9510396].

\bibitem{BramonGosdzinsky}
Bramon, A., Gosdzinsky, P. and Tortosa, S.,
Phys.\ Lett.\ B {\bf 377}, 140 (1996)
[hep-ph/9603357].

\bibitem{Neufeldetal}
D'Ambrosio, G., Ecker, G., Isidori, G. and Neufeld, H.,
Phys.\ Lett.\ B {\bf 466}, 337 (1999)
[hep-ph/9905420].

\bibitem{dashen}
Dashen, R. F.,
Phys.\ Rev.\  {\bf 183}, 1245 (1969).

\bibitem{GLeta}
Gasser, J. and Leutwyler, H.,
Nucl.\ Phys.\ B {\bf 250}, 539 (1985).

\bibitem{ABT1}
Amoros, G., Bijnens, J. and Talavera, P.,
Phys.\ Lett.\ B {\bf 480}, 71 (2000)
[hep-ph/9912398];\\
Nucl.\ Phys.\ B {\bf 585}, 293 (2000)
[Erratum-ibid.\ B {\bf 598}, 665 (2000)]
[hep-ph/0003258].

\bibitem{Truong}
Roiesnel, C. and Truong, T.,
Nucl.\ Phys.\ B {\bf 187}, 293 (1981).

\bibitem{KhuriTreiman}
Khuri, N. N. and Treiman, S. B.,
       Phys. Rev. {\bf 119}, 1115  (1960);\\
Kacser, C., Phys. Rev. {\bf 132}, 2712-2721 (1963)

\bibitem{KWW}
Kambor, J., Wiesendanger, C. and Wyler, D.,
Nucl.\ Phys.\ B {\bf 465}, 215 (1996)
[hep-ph/9509374].

\bibitem{KnechtStern}
Knecht, M., Moussallam, B., Stern, J. and Fuchs, N. H.,
Nucl.\ Phys.\ B {\bf 471}, 445 (1996)
[hep-ph/9512404].

\bibitem{AnisovichLeutwyler}
Anisovich, A. V. and Leutwyler, H.,
Phys.\ Lett.\ B {\bf 375}, 335 (1996)
[hep-ph/9601237].

\bibitem{Anisovich}
Anisovich, A. V.,
Phys.\ Atom.\ Nucl.\  {\bf 58}, 1383 (1995)
[Yad.\ Fiz.\  {\bf 58N8}, 1467 (1995)].

\bibitem{CB2} 
Tippens, W. B. {\it et al.}  [Crystal Ball Collaboration],
Phys.\ Rev.\ Lett.\  {\bf 87}, 192001 (2001).
Eq. (1c) in this paper contains a misprint: The $\sqrt3$ in the
definition of $y$ should be replaced by 3.

\bibitem{Layter}
Layter, J. G. {\it et al.},
Phys.\ Rev.\ D {\bf 7} (1973) 2565.

\bibitem{Gormley}
Gormley, M. {\it et al.},
Phys.\ Rev.\ D {\bf 2}, 501 (1970).

\bibitem{Amsler}
Amsler, C. {\it et al.}  [Crystal Barrel Collaboration],
Phys.\ Lett.\ B {\bf 346}, 203 (1995).

\bibitem{Amsler2}
Abele, A. {\it et al.}  [Crystal Barrel Collaboration],
Phys.\ Lett.\ B {\bf 417}, 197 (1998).

\bibitem{Alde}
Alde, D. {\it et al.}  [Serpukhov-Brussels-Annecy(LAPP) Collaboration],
Z.\ Phys.\ C {\bf 25}, 225 (1984)
[Yad.\ Fiz.\  {\bf 40}, 1447 (1984)].

\bibitem{Amsler3}
Abele, A. {\it et al.}  [Crystal Barrel Collaboration],
Phys.\ Lett.\ B {\bf 417}, 193 (1998).

\bibitem{Nefkens} Nefkens, B. and Price, J., these proceedings.

\bibitem{BijnensPersson}
Bijnens, J. and Persson, F.,
``Effects of different form-factors in meson photon photon transitions
 and the muon anomalous magnetic moment,''
 [hep-ph/0106130]. (Master's thesis of F.~Persson.)

\bibitem{formasympt}
Novikov, V.A., M.~A.~Shifman, M. A., Vainshtein, A. I., 
Voloshin, M. B. and Zakharov, V. I.,
Nucl.\ Phys.\ B {\bf 237}, 525 (1984);\\
Nesterenko, V. A. and A.~V.~Radyushkin, A. V.,
Sov.\ J.\ Nucl.\ Phys.\  {\bf 38}, 284 (1983)
[Yad.\ Fiz.\  {\bf 38}, 476 (1983)].

\bibitem{form4try}
Knecht, M., Peris, S., Perrottet, M. and de Rafael, E.,
Phys.\ Rev.\ Lett.\  {\bf 83}, 5230 (1999)
[hep-ph/9908283].

\bibitem{form4extra}
Knecht, M. and Nyffeler, A.,
Eur.\ Phys.\ J.\ C {\bf 21}, 659 (2001)
[hep-ph/0106034].

\bibitem{g-2}
Knecht, M. and Nyffeler, A.,
[hep-ph/0111058];\\
Knecht, M., Nyffeler, A., Perrottet, M. and de Rafael, E.,
 [hep-ph/0111059];\\
Hayakawa, M. and Kinoshita, T.,
Phys.\ Rev.\ D {\bf 57}, 465 (1998)
[hep-ph/9708227];\\
Bijnens, J., Pallante, E. and Prades, J.,
Phys.\ Rev.\ Lett.\  {\bf 75}, 1447 (1995)
[Erratum-ibid.\  {\bf 75}, 3781 (1995)]
[hep-ph/9505251];\\
Bijnens, J., Pallante, E. and Prades, J.,
Nucl.\ Phys.\ B {\bf 474}, 379 (1996)
[hep-ph/9511388];\\
Bijnens, J., Pallante, E. and J.~Prades, J.,
[hep-ph/0112255];\\
Hayakawa, M. and Kinoshita, T.,
[hep-ph/0112102].

\bibitem{ABBC}
Ametller, Ll., Bijnens, J., Bramon, A. and Cornet, F.,
Phys.\ Lett.\ B {\bf 276}, 185 (1992).

\bibitem{epggexp}
Achasov, M. N. {\it et al.},
Nucl.\ Phys.\ B {\bf 600}, 3 (2001)
[hep-ex/0101043].

\bibitem{epggenjl}
Nemoto, Y., Oka, M. and Takizawa, M.,
Phys.\ Rev.\ D {\bf 54}, 6777 (1996)
[hep-ph/9602253];\\
Bijnens, J., Fayyazuddin, A. and Prades, J.,
Phys.\ Lett.\ B {\bf 379}, 209 (1996)
[hep-ph/9512374];\\
Bel'kov, A. A., Lanyov, A. V. and Scherer, S.,
J.\ Phys.\ G {\bf 22}, 1383 (1996)
[hep-ph/9506406];\\
Bellucci, S. and Bruno, C,,
Nucl.\ Phys.\ B {\bf 452}, 626 (1995)
[arXiv:hep-ph/9502243].

\bibitem{Koetal}
Ko, P.,
Phys.\ Lett.\ B {\bf 349}, 555 (1995)
[hep-ph/9503253];\\
Phys.\ Rev.\ D {\bf 47}, 3933 (1993).

\bibitem{quarkmodel}
Ng, J. N. and Peters, D. J.,
Phys.\ Rev.\ D {\bf 47}, 4939 (1993).
\end{thebibliography}
\end{document}